\documentclass[prb,aps,twocolumn,showpacs,superscriptaddress]{revtex4}
\usepackage{graphicx}
\usepackage{amsmath}

\begin{document}
	
\title{Fate of partial order on the trillium and distorted windmill lattices}

\author{Sergei V. Isakov}
\affiliation{Institute for Theoretical Physics, ETH Zurich, CH-8093 Zurich, Switzerland}
\author{John M. Hopkinson}
\affiliation{Brandon University, Brandon, Manitoba, Canada, R7A 6A9}
\author{Hae-Young Kee}
\affiliation{University of Toronto, Toronto, Ontario, Canada, M5S 1A7}
\affiliation{Korea Institute for Advanced Study, Seoul 130-722, Korea}


\begin{abstract}
The classical Heisenberg model on the trillium and distorted windmill
lattices exhibits a degenerate ground state within large-$N$ theory,
where the degenerate wavevectors form a surface and line, in 3-dimensional space, respectively.   
We name such states partially ordered to represent the existence of
long-range order along the direction normal to these degenerate manifolds. 
We investigate the effects of thermal fluctuations using Monte Carlo (MC) methods,
and find a first order transition to a magnetically ordered state for both cases.
We further show that the ordering on the distorted windmill lattice
is due to order by disorder, while the ground state of the trillium lattice is unique.
Despite these different routes to the realization of low temperature ordered phases,
the static structure factors obtained by large-$N$ theory and MC simulations for each lattice 
show quantitative agreement in the cooperative paramagnetic regime at finite temperatures.
This suggests that a remnant of the characteristic angle-dependent spin correlations  of partial order remains 
above the transition temperatures for both lattices. 
The possible relevance of these results to $\beta$-Mn, CeIrSi, and MnSi is discussed.
\end{abstract} 

\maketitle

\section{Introduction}

\begin{figure}[here]
\includegraphics[scale=0.43]{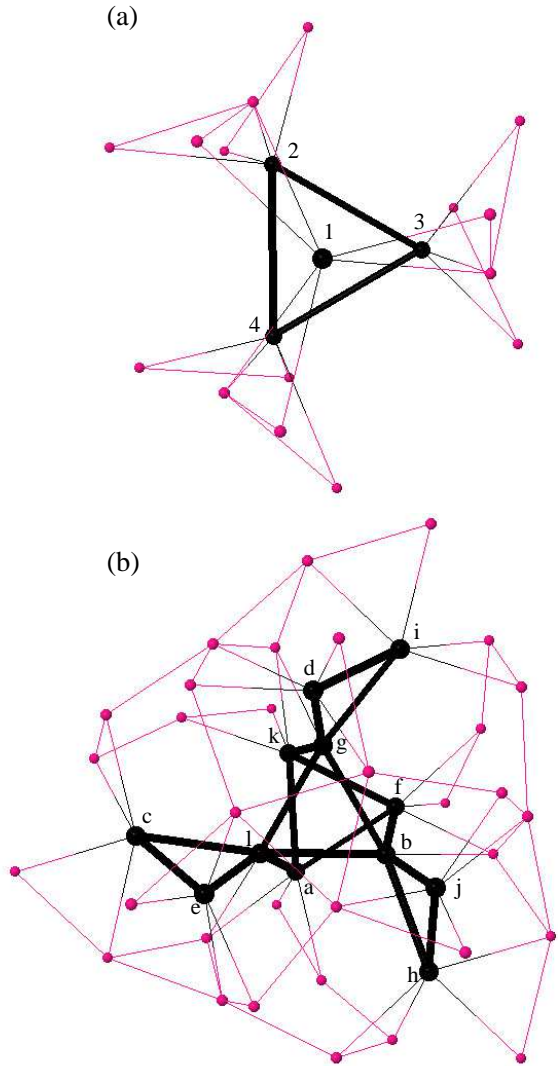}
\caption{\label{figure1}  (Color online) Magnetic sites of (a) the trillium lattice and (b) the distorted windmill lattice.  In both cases three corner-shared equilateral triangles meet at each spin.  The unit cell of the trillium lattice is only 4 sites and the next smallest spin loop is 5 sided.  The unit cell of the distorted windmill lattice is 12 sites and the smallest spin loop is 4 sided. Atoms within the first unit cell are depicted\cite{jmol} in boldface (black) and labeled ((b) as in Table I).  }
\end{figure}

Geometrically frustrated (GF) magnetic systems offer a rich avenue to the search 
for emergent or collective phases of matter.
Frustrated magnetism arises when spins between nearby magnetic sites cannot 
form a unique alignment to minimize their magnetic interactions,
and leads to the possibility of a macroscopically degenerate classical ground state. 
Materials with antiferromagnetic nearest neighbor interactions between local moments 
lying at the sites of corner-sharing tetrahedra or triangles comprise a considerable 
fraction of the GF magnetic systems known to date. 

Most magnetic materials select a particular ordering wavevector, representative of their magnetic structure, upon the onset of magnetic order.  In contrast,
in frustrated magnets, it is not uncommon to find within the mean field
approximation a set of degenerate continuously connected ``ordering wavevectors" 
which span the three-dimensional (3-D) space.  
For example, the disordered ground state of the classical Heisenberg model on the pyrochlore lattice \cite{reimersmc, moessenrchalker}
features such a wavevector manifold.  
Partially ordered magnets lie between these two extremes, forming degenerate wavevector (2-D) surfaces or (1-D) lines in a 3-D space. 
The name of ``partially ordered'' comes from the fact that
there is a long-range order along the direction normal to this degenerate manifold. 
In other words, in such systems, to very low temperatures, a system may appear 
disordered/ordered to measurements probing parallel/perpendicular directions to this line or surface.  
A growing number of frustrated spin models have been shown to exhibit such a partial order.\cite{hopkinson,canals,bergman,unpublishedus}
Experimental evidence for partial order has been observed in single-crystal neutron scattering measurements 
of the correlated metals, MnSi{\cite{Pflei}} under pressure, and CeCu$_{5.9}$Au$_{0.1}$\cite{Schroder}, although in these materials it is unlikely to arise from considerations as simple as here considered.

Our interest in this problem arose from the realization that partial order has been found using large-$N$ theory on both the trillium and distorted windmill lattices. 
The key question we address in this paper is whether these partially ordered states can be captured
beyond the large-$N$ theory.\cite{noteref3} If not, to what extent can the results of large-$N$ theory
be regained due to thermal fluctuations?
We answer this question by carrying out large scale classical Monte Carlo simulations of the AF Heisenberg 
model on the trillium and distorted windmill lattices.  

Large-$N$ theory and MC simulations of the classical Heisenberg model give consistent results for both highly frustrated systems (ex: on the pyrochlore lattice a disordered ground state is found), and for unfrustrated lattices, (ex: choosing the correct ordering wavevector).\cite{note1a}  Despite this trend, the finding of a partially ordered ground state in large-$N$ theory does {\it{not}} necessarily translate into a partially ordered ground state within MC simulations.
We find that the classical Heisenberg model on both the trillium and distorted windmill lattices shows a first order phase transition to a coplanar magnetically ordered state featuring neighboring spins rotated by 120$^\circ$.  
We further show that this model on the trillium lattice does not have a macroscopic ground state degeneracy.
The chosen 120$^\circ$ coplanar state is a unique ground state, and the partial order is an artificial effect of the spherical approximation
constraint\cite{berlin} used in the large-$N$ theory\cite{hopkinson}.
On the other hand,  the selection of a particular coplanar ordering on the distorted windmill lattice
proceeds due to an order by disorder mechanism.

In the cooperative paramagnetic state (for an intermediate temperature regime $T_c<T<\theta_{CW}$ extending an order of 
magnitude below the intercept of a Curie-Weiss fit, $\theta_{CW}$),  
we find that the large-$N$  description remains quantitatively valid for both lattices, despite the inability of large-$N$ theory to capture the true nature of the classical ground state.  
That is, calculations of the angle-resolved static structure factor by large-$N$  and MC techniques show 
at most a 5\% difference.  We will henceforth refer to this temperature window as the cooperative 
paramagnetic regime as it features reasonably strong but short range spin correlations, finite temperature remnants of an avoided partial order. 

In the next section, we present the structures of the two lattices, and discuss the approaches used
to study the magnetic properties of the classical Heisenberg model on these lattices.
We show the results of MC simulations on the trillium and distorted windmill lattices in Sec. III and IV, respectively.
In particular, we focus on the questions addressed above:
``is there a transition to a magnetically  ordered state
at finite temperatures'' and 
``if so, do we find remnants of the partial order above the transition temperature as a result of thermal fluctuations''.
We also address the mechanism of ordering in these lattices after an enumeration of their ground states.
The possible relevance of these results to real materials and the conclusion are discussed in Sec. V.  A heuristic derivation of the degeneracy of the model on the distorted windmill lattice is presented in Appendix A.

\section{Model, lattice structures, and approaches}

We study the classical $O(N)$ model on the trillium and distorted windmill
lattices given by the following Hamiltonian
\begin{equation}
  H=J \sum_{\langle ij \rangle} \mathbf{S}_i \cdot \mathbf{S}_j,
\end{equation}
where $J>0$ is the antiferromagnetic exchange coupling constant, the sum runs over the nearest
neighbors only, and $\mathbf{S}=(S^1,\ldots,S^N)$ is an $N$-component
classical spin.

It was shown in Ref.~\onlinecite{hopkinson} that
the magnetic lattice of MnSi forms a three-dimensional network of corner-sharing equilateral triangles 
with the cubic P2$_1$3 symmetry shown in Fig.~1 (a), which we have named the trillium lattice.  
The trillium lattice is common to many systems including the antiferromagnetically correlated Ce local moments of CeIrSi.
The magnetic lattice of $\beta$-Mn, the distorted windmill lattice (P4$_1$32 symmetry), 
bears a remarkable qualitative resemblance to the trillium lattice as shown in Fig.~1 (b). In particular, both lattice structures feature three corner-shared equilateral triangles joined at a common site.  
The coordinates of  each site within a unit cell for both the trillium lattice \cite{hopkinson} and the distorted windmill lattice\cite{nakamura1} have been previously presented.  For completeness, these are listed in Table \ref{TableI}

\begin{table}[hbtp]
\begin{tabular}{|l|l|}
\hline
label&coordinates\\
\hline
1&$(u,u,u)$\\
\hline
2&$(\frac{1}{2}+u,\frac{1}{2}-u,1-u)$\\
\hline
3&$(1-u,\frac{1}{2}+u,\frac{1}{2}-u)$\\
\hline
4&$(\frac{1}{2}-u,1-u,\frac{1}{2}+u)$\\
\hline
$a$&$(\frac{1}{8},y,\frac{1}{4}+y)$\\
\hline
$b$&$(\frac{5}{8},\frac{1}{2}-y,\frac{3}{4}-y)$\\
\hline
$c$&$(\frac{1}{4}-y,\frac{7}{8},\frac{1}{2}+y)$\\
\hline
$d$&$(1-y,\frac{3}{4}+y,\frac{3}{8})$\\
\hline
$e$&$(\frac{3}{8},1-y,\frac{3}{4}+y)$\\
\hline
$f$&$(\frac{1}{4}+y,\frac{1}{8},y)$\\
\hline
$g$&$(\frac{3}{4}-y,\frac{5}{8},\frac{1}{2}-y)$\\
\hline
$h$&$(\frac{1}{2}+y,\frac{1}{4}-y,\frac{7}{8})$\\
\hline
$i$&$(\frac{7}{8},\frac{1}{2}+y,\frac{1}{4}-y)$\\
\hline
$j$&$(\frac{3}{4}+y,\frac{3}{8},1-y)$\\
\hline
$k$&$(y,\frac{1}{4}+y,\frac{1}{8})$\\
\hline
$l$&$(\frac{1}{2}-y,\frac{3}{4}-y,\frac{5}{8})$\\
\hline
\end{tabular}
\caption{\label{TableI}
The coordinates of the trillium and distorted windmill lattice sites in the first unit cell as shown in Fig. 1, are expressed in terms of lattice parameters $u$, and $y$ respectively, where we set the corresponding lattice constants of the cubic lattices to unity.
 }
\end{table}

To carry out a quantitative comparison between large-$N$ theory and $N=3$ Monte Carlo results,
we have carried out large scale classical Monte Carlo simulations of the $N$=3 Heisenberg model on the trillium and distorted windmill lattices for lattices with $n_s=n\times L\times L\times L$ spins, where $n$ is the number of sites (or spins) in the unit cell and $L$ is the number of spins along each dimension of a cube.  On the trillium lattice ($n$=4), we have considered $L=\{6,9,12,18\}$. 
On the distorted windmill lattice ($n=12$), we have considered $L=\{6,8,12\}$.
The standard Metropolis algorithm has been used in which we attempt to update
a spin within a small angular range $\delta$ around its original direction.
We choose $\delta$ in such a way that around 50\% of attempted spin updates
are accepted. Starting with a random configuration, we usually perform
$2 \cdot 10^5$ Monte Carlo steps (MCS) for equilibration and $10^6$ MCS for
measurements with one MCS consisting of $\sim n_s/T$ single spin updates,
where $T$ is the temperature.

Within large$-N$ theory, one writes the partition function for the $N$-component spin model\cite{stanley} and solves for the spin-spin correlations \cite{garcan, isakov, hopkinson} in the limit $N\rightarrow\infty$.
That is we rewrite the Hamiltonian as
\begin{equation}
H = {T \over 2} \sum_{i,j} M_{ij} \mathbf{S}_i \cdot \mathbf{S}_j,
\end{equation}
where $M_{ij}$ is the interaction matrix and the spins are subject to the constraint
$\mathbf{S}_i^2=N$. The corresponding partition function is given by
\begin{equation}
Z = \int {\cal D} \phi  {\cal D} \lambda \ e^{-S({\phi},\lambda)}
\label{eq:partfunc}
\end{equation}
with the action 
\begin{equation}
S({\phi},\lambda) = \sum_{i,j}  \left [ {1 \over 2}
M_{ij} {\phi}_i \cdot  {\phi}_j +  {\lambda_i \over 2} \delta_{ij}
({\phi}_i \cdot {\phi}_i - N) \right ],
\label{eq:action}
\end{equation}
where ${\phi}_i=(\phi^1_i, ..., \phi^N_i)$ is an $N$-component real 
vector field and $\lambda_i$ the Lagrange multiplier for the
constraint ${\phi}_i \cdot {\phi}_i = N$. To proceed, we take
the $N \rightarrow \infty$ limit and set a uniform $\lambda_i = \lambda_0$. 
The locations $i = (l,\mu)$ of spins can be labeled by 
those of the cubic unit cell $l = 1,\ldots, n_c$ and the lattice
sites  $\mu=1,\ldots,n$ within the unit cell ($n_c=L\times L\times L$ is
the total number of the unit cells in the lattice).
The Fourier transform with respect to the positions of the unit cells leads to
\begin{equation}
S({\phi}) = \sum_{\bf q} \sum_{\mu,\nu} {1 \over 2}
A^{\mu \nu}_{\bf q} {\phi}_{{\bf q},\mu} \cdot {\phi}_{{\bf q},\nu}
\end{equation}
with $A^{\mu \nu}_{\bf q} = M^{\mu \nu}_{\bf q} + \delta_{\mu \nu} \lambda_0$.
Performing Gaussian integrations over the ${\phi}$ fields in
Eq.~\ref{eq:partfunc}, one finds that
$\lambda_0$ is determined by the saddle point equation,
\begin{equation}
n n_c=\sum_q \sum_{\rho=1}^n\frac{1}{\beta \epsilon_q^\rho+\lambda_0},
\end{equation}
where $\beta\epsilon_q^\rho$ are the eigenvalues of the $n \times n$ 
interaction matrix $M^{\mu \nu}_{\bf q}$ (which has been shifted such that Min$(\epsilon_q^\rho)=0)$  and $\beta$ the inverse temperature.
Note that the sum over $q$ is carried out for finite size periodic lattices to effectively compare with the Monte Carlo results.
Plugging $\lambda_0$ back into Eqs.~\ref{eq:partfunc} and \ref{eq:action},
we readily deduce the spin-spin correlation functions. For example,
these can be found by calculating the second derivative of $Z$ with respect to an auxiliary field which couples to the spin and can be added to Eq. \ref{eq:action}. 
The static structure factor is found to be
\begin{equation}
  S(\mathbf{q}) \propto
    \sum_{\kappa,\kappa'=1}^n\langle S_\mathbf{q}^{\kappa'}S_{-\mathbf{q}}^{\kappa}\rangle
    =\sum_{\kappa,\kappa',\rho=1}^n
      \frac{U_{\kappa'\rho}U^*_{\kappa\rho}}{\beta\epsilon_q^\rho+\lambda_0},
\end{equation}
where $U$ is the matrix that diagonalizes the interaction matrix
$M^{\mu \nu}_{\bf q}$, $\kappa$ and $\rho$ are the sublattice indices.
Note that, for simplicity, we will normalize spins to $\sqrt{N}$ in the
large-$N$ theory.\cite{prv} To compare with MC results ($N=3$), one needs to fix the
energy scale. Here we rescale $J\rightarrow J/3$ in $\epsilon_{q}^{\rho}$.

\section{The trillium lattice}

Within large-$N$ theory, we reported\cite{hopkinson} that the classical antiferromagnetic (AF) Heisenberg model on the trillium lattice has a partially ordered ground state, with a surface of degenerate wavevectors following
\begin{equation}
\cos^2\left(\frac{q_x}{2}\right)+\cos^2\left(\frac{q_y}{2}\right)
  +\cos^2\left(\frac{q_z}{2}\right)=\frac{9}{4},\label{equation7}
\end{equation}
where we set the lattice constant $a=1$.  
However, in Ref. \onlinecite{hopkinson} we relaxed the hard spin constraint of $\mathbf{S}_i^2=1$ to a soft constraint, $\sum^{n}_{i=1} \mathbf{S}_i^2 = n$, where $n$ is the number of sites in the unit cell.
Using energy minimization on spin clusters of classical spins (with $N$=3), we effectively imposed the hard spin constraint show  
that the lowest energy state actually exhibits a coplanar magnetic order with wavevector ($\frac{2\pi}{3}$,0,0),\cite{andothers}
.
We speculated that the soft constraint of conventional large-$N$ theory is crucial to the realization of partial order as $T \rightarrow 0$\cite{hopkinson}.

\begin{figure}
\includegraphics[width=3in]{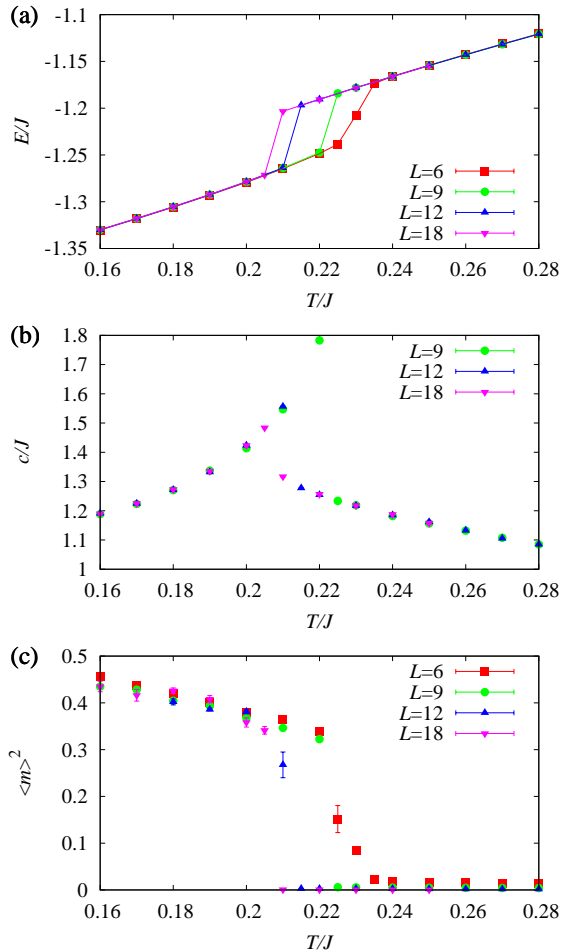} 
\caption{\label{figure2b} (Color online) Average energy per spin (a), heat capacity (b) and magnetic order parameter (c) vs. temperature for the Heisenberg model on the trillium lattice. These results are independent of the lattice parameter $u$ which determines the position of the sites in a cubic unit cell. In this and all other plots, error bars are smaller
than the symbol size if not visible.}
\end{figure}

To find out whether a finite temperature transition to a magnetically ordered state occurs,
we have carried out large scale MC simulations. The energy as a function of temperature, as seen in 
Fig.~\ref{figure2b} (a), exhibits a smooth behavior before an abrupt jump, which indicates a first order transition to an ordered state.
The heat capacity as a function of temperature, shown in Fig.~\ref{figure2b} (b),  clearly shows a peak corresponding to the onset of magnetic order on the trillium lattice. 
Because of strong hysteresis effects, the location of this jump appears to
scale to lower temperatures as the size of the system increases. We have
not attempted to determine the precise location of this strongly first order
transition. Our best estimate for the transition temperature is $T_c=0.21(1)J$.
In Fig.~\ref{figure2b} (c), we plot the magnetic order parameter of this transition, 
which is defined as the structure factor at the ordering wavevector,\cite{specifically} $\mathbf{Q} = (\frac{2 \pi}{3}, 0,0)$:
\begin{equation}
  \langle m \rangle^2={S(\mathbf{q}=\mathbf{Q}) \over n_s}.
\end{equation}
We see that this order parameter exhibits a sharp onset at the transition temperature as would be expected for a first order phase transition.

It was anticipated that the MC would find a transition to a 120$^\circ$ coplanar magnetic state with
the wavevector ($\frac{2\pi}{3},0,0$), where the spins on each triangle form at 120$^\circ$ to each other,
since it was shown (by minimization) that this is the lowest energy state. Thus this ground state does not have a macroscopic degeneracy\cite{hopkinson}
and the observed transition is not expected to result from an order by disorder mechanism.
An intuitive way to understand the uniqueness of this ground state is to start with a coplanar state 
with spins labeled $\alpha$, $\beta$ and $\gamma$ where the letters 
 $\alpha$, $\beta$ and $\gamma$ denote 120$^\circ$ rotated spins on every triangle. 
It is then natural to ask whether
one can generate degenerate states by rotating $\beta$ and $\gamma$ spins around the axis of spin $\alpha$. 
The high connectivity of this lattice in comparison with the kagome and hyperkagome lattices\cite{lawler} where such degeneracies naturally arise, prevents any such rotations which are not consistent with the crystal symmetries of the lattice.
Therefore, the rotation of $\beta$ and $\gamma$ spins around $\alpha$ spin axis cannot generate distinctly different states.  
This argument itself is not sufficient to prove that
the ground state is unique because, in principle, there might be other
ground states that are not connected to the $(\frac{2\pi}{3},0,0)$ state
by simple spin rotations.  However, our previous minimization showed that the state with
the wavevector $(\frac{2\pi}{3},0,0)$ is the lowest energy state. Combining our current MC
result, we conclude that the ground state of Heisenberg model on the trillium lattice is unique.

In the disordered state at temperatures above this ordering transition, however, the spins fluctuate, and one might expect to recover features found by large-$N$ theory with the soft constraint.
Now the question is
{\it{to what extent is the partial order of large-$N$  theory recovered due to temperature fluctuations?}}

\begin{figure}
\includegraphics[scale=0.4]{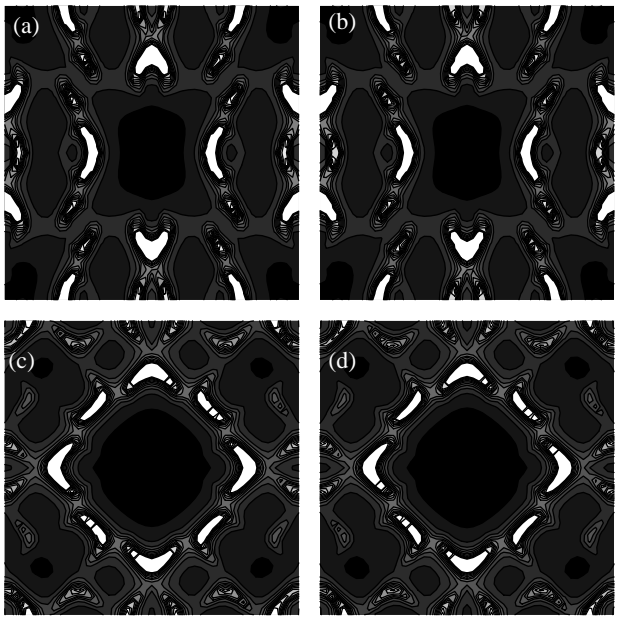}
\caption{\label{figure3}   Contour plots of the intensity of the structure factor in the hhk ((a) and (b)) and hk0 ((c) and (d)) planes for the trillium lattice in the cooperative paramagnetic regime (for $u=0.138$, $L=12$ and $T=0.25J$) show prominant features near these surfaces. Classical Monte Carlo ((a) and (c)) agrees well with large-$N$ results ((b) and (d)). The maximal intensity is shown in white, and the axes run from $-4\pi$ to $4\pi$ along k(001) (vertical), h(110) (horizontal), k(010) (vertical), h(100)(horizontal). Notice the prominent features near the zero energy surfaces given by Eq. \ref{equation7} of the large-$N$ theory. }
\end{figure}
\begin{figure}
\includegraphics[width=3in]{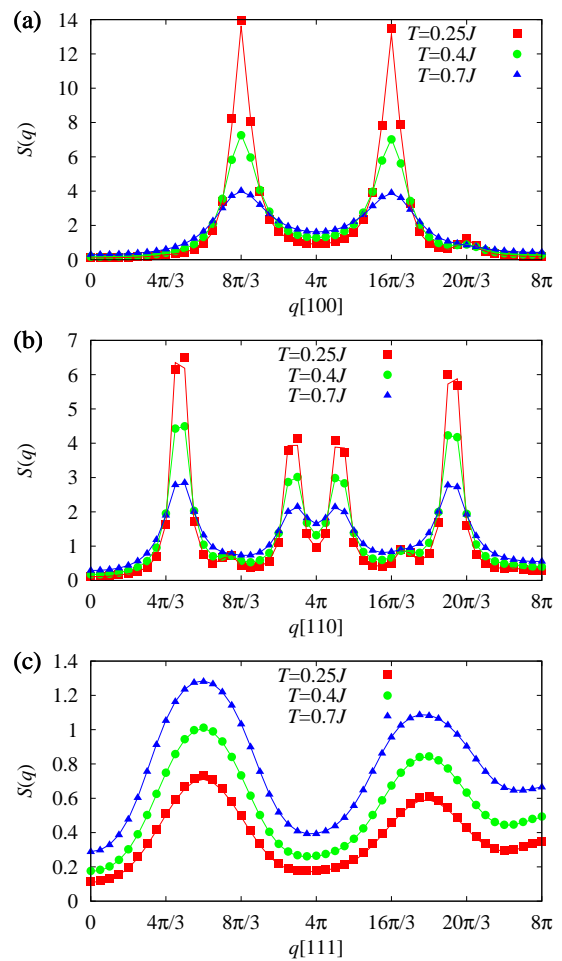}
\caption{\label{figure5} (Color online) A quantitative comparison of the angle-dependent structure factor is shown along three high symmetry directions between the Monte Carlo (Heisenberg model) and large-$N$ results for classical spins in the cooperative paramagnetic phase on the trillium lattice ($u=0.138,L=12$). The solid line is the large-$N$  result, while the symbols are from MC simulations.}
\end{figure}

Magnetic susceptibility (not shown) has been calculated within both approaches and shows good agreement until very close to the ordering temperature $T_c\approx0.21J$, well below the Curie-Weiss temperature $\Theta_\text{CW}=2J$, and in the zero temperature limit.
Qualitative comparisons between the large-$N$ theory and Monte Carlo results for the structure factor are shown in Fig.~\ref{figure3} in the cooperative paramagnetic phase at temperatures slightly above the onset of magnetic order.  We see excellent qualitative agreement between these two approaches.  Within the cooperative paramagnetic window we see that the static structure factor on the trillium lattice peaks around the surface of a sphere-like shape determined by Eq. \ref{equation7}, although part of this sphere is 
obliterated by geometric (rather than energetic) effects. 

Within both approaches the geometric factor (that is cancellations between spin contributions within the unit cell) imposes that there is very little weight within the first Brillouin zone.  In Fig.~\ref{figure5} a quantitative comparison along several high symmetry directions between these two approaches is presented.  There is extremely good agreement at all wavevectors. 
Note that the cooperative paramagnetic regime is smoothly connected to the partial order within large-$N$ approaches.
Therefore excellent agreement between the angle dependent spin-spin correlation functions implies  that remnants of partial order would be expected to appear at finite temperatures above the transition temperature as a result of thermal fluctuations.

\begin{figure}
\includegraphics[scale=0.67]{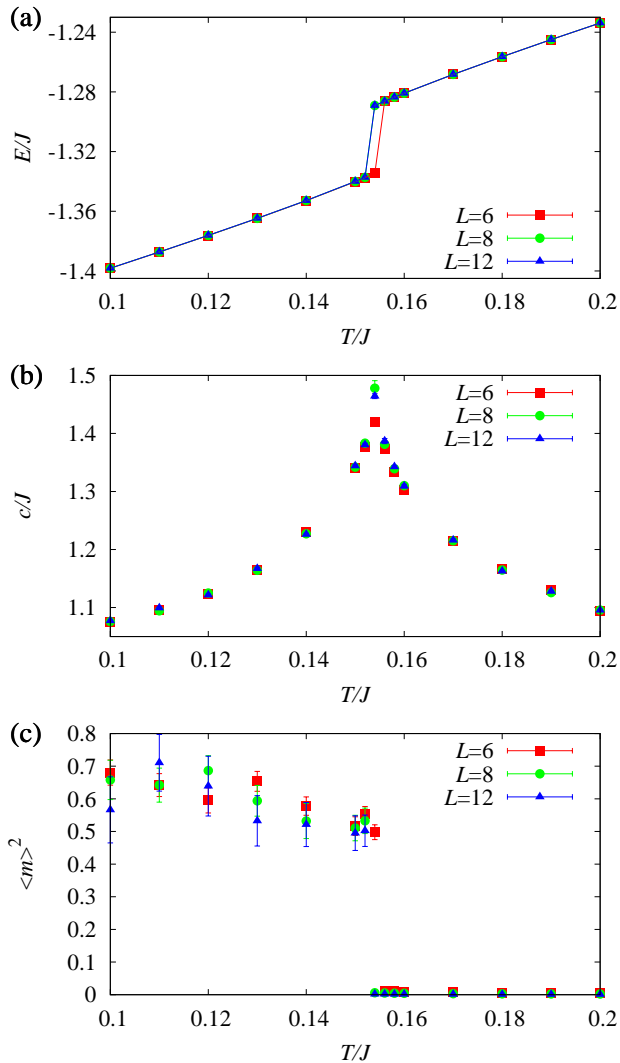} 
\caption{\label{figure2} (Color online) Average energy per spin (a), heat capacity (b), and magnetic order parameter (c) vs. temperature for the Heisenberg model on the idealized $\beta$-Mn lattice.}
\end{figure}

\section{Distorted windmill lattice}

Mean field calculations\cite{canals} on the distorted windmill lattice also revealed a partially ordered ground state, with a line of degenerate wavevectors along the $(qqq)$ direction.  In this paper we show that Monte Carlo simulations of the AF Heisenberg model on this lattice find an ordered ground state as on the trillium lattice.  However, the physical origin of the two phase transitions differs. While the ordering in the trillium lattice can be understood by minimization, being the result of energetically different states, the ordering in the distorted windmill lattice cannot.  Rather, it must proceed via an order by disorder mechanism.
As on the trillium lattice it is interesting to ask whether the partial ordering features obtained by large-$N$ (or mean field) theory can be found at finite T, where one might expect the spins to fluctuate strongly.\cite{note}

Again, on the distorted windmill lattice, MC simulations find
a jump in the energy as a function of temperature indicative of a first order phase transition. Nonetheless, this transition is qualitatively different from that seen on the trillium lattice, as it does not show strong hysteresis
effects and much variation with system size.  This is reflected in a sharp transition from one smooth energy vs. temperature curve to another at the transition temperature as seen in Fig.~\ref{figure2} (a).
The heat capacity as a function of temperature, shown in Fig.~\ref{figure2} (b), clearly shows a peak corresponding to the onset of magnetic order.
The variation of the order parameter shows an abrupt onset at the transition temperature.  Here the order parameter has been defined in terms of the structure factor at the ordering wave vector,\cite{specifically}
$\mathbf{Q}_0 = (0,0,0)$ as
\begin{equation}
  \langle m \rangle^2={S(\mathbf{q}=\mathbf{Q}_0) \over n_s}.
\end{equation}
As in the case of the trillium lattice, we have not attempted to determine
the precise location of this first order transition. Our best estimate for
the transition temperature is $T_c=0.155(5)J$.

To understand this spin ordering we have carried out an energetic minimization of spin clusters with $N=3$.
The minimization of spin clusters on the distorted windmill lattice with $qqq$ symmetry is found to admit only two types of magnetic spin substructures: one of which can repeat as it is from one unit cell to the next, and another which undergoes a 120$^\circ$ spin rotation in progressing by one unit cell along the ($111$) direction. Thus one has topologically distinct ground state sectors, unit cell structures which are shown in Table \ref{TableII} and labeled $s_{q=0}$ and $s_{q=2\pi/3}$ respectively,
where the ordering wavevector in $s_{q=2\pi/3}$ is ${\bf Q}_{sp}=(\frac{2\pi}{3},\frac{2\pi}{3},\frac{2\pi}{3})$.  
Curiously, as each spin structure features coplanar 120$^\circ$ rotated spins on each triangle, generic ground state spin structures are found to interchange between these two orderings.  The only constraint arising for a $s_{q=2\pi/3}$ spin structure is that the spin structure must complete three times an integral number of $q=\frac{2\pi}{3}$ unit cells within its boundaries.   
There are a macroscopic number of degenerate ground states which can be generated by this mixing of
these two states, all of which exhibit a coplanar order.  
This tendency to form a coplanar spin structure is reminiscent of the nematically ordered state recently found for the classical Heisenberg model on the hyperkagome lattice.\cite{usprl} We estimate that the degeneracy of the lattice grows exponentially in the linear lattice size, L, roughly as $e^{0.69L}/6$, 
as shown in Appendix A.

\begin{figure}
\includegraphics[width=3in]{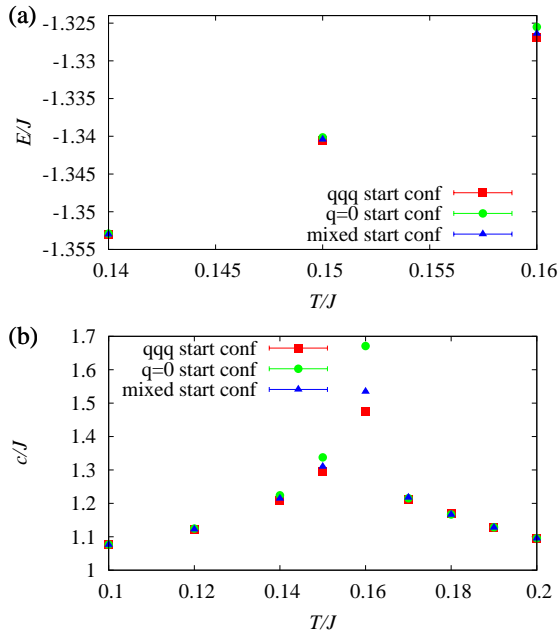}
\caption{\label{energy_0vssp} (Color online) For lattice size $L=6$, (a) Energy and (b) Heat capacity vs. temperature 
for the AF Heisenberg model on the distorted windmill lattice with different initial spin configurations.
The filled squares, circles, and triangles represent the $\mathbf{Q}_\text{0}$,
$\mathbf{Q}_\text{sp}$, and $\mathbf{Q}_\text{mix}$ spin configurations, respectively. The system remains in its initial configuration below the
transition. Note that starting with any random spin configurations leads to the $\mathbf{Q}_\text{0}$ state.
(a) shows that the $\mathbf{Q}_\text{sp}$ and $\mathbf{Q}_\text{mix}$ states have the same energy as the $\mathbf{Q}_\text{0}$ state as $T \rightarrow 0$,
and have lower energy than the $\mathbf{Q}_\text{0}$ state just below the transition temperature. However,
(b) indicates that the $\mathbf{Q}_\text{sp}$ state has lower entropy below the transition than the $\mathbf{Q}_\text{mix}$ state and that state in turn has lower
entropy than $\mathbf{Q}_\text{0}$ state.
 This is evidence that the selection of the $\mathbf{Q}_\text{0}$ state is of entropic origin, i.e., order
by disorder.}
\end{figure}

\begin{table}[hbtp]
\begin{tabular}{|l|l|l|l|l|l|}
\hline
label&near. neighb. 1&near. neighb. 2&$s_{q=0}$&$s_{q=2\pi/3}$\\
\hline
$a$&$c_{-\hat{y}},d_{-\hat{x}-\hat{y}},f,k$&$j_{-\hat{x}},l$&$\alpha$&$\alpha$\\
\hline
$b$&$g,h,j,l$&$d_{-\hat{y}},f$&$\alpha$&$\alpha$\\
\hline
$c$&$a_{+\hat{y}},d_{-\hat{x}},e,l$&$h_{-\hat{x}+\hat{y}},i_{-\hat{x}+\hat{z}}$&$\beta$&$\alpha$\\
\hline
$d$&$a_{+\hat{x}+\hat{y}},c_{+\hat{x}},g,i$&$b_{+\hat{y}},f_{+\hat{y}}$&$\gamma$&$\alpha$\\
\hline
$e$&$c,f_{+\hat{y}+\hat{z}},h_{+\hat{y}},l$&$g_{+\hat{z}},k_{+\hat{z}}$&$\alpha$&$\beta$\\
\hline
$f$&$a,e_{-\hat{y}-\hat{z}},h_{-\hat{z}},k$&$b,d_{-\hat{y}}$&$\beta$&$\beta$\\
\hline
$g$&$b,d,i,l$&$e_{-\hat{z}},k$&$\beta$&$\beta$\\
\hline
$h$&$b,e_{-\hat{y}},f_{+\hat{z}},j$&$c_{+\hat{x}-\hat{y}},i_{-\hat{y}+\hat{z}}$&$\gamma$&$\beta$\\
\hline
$i$&$d,g,j_{-\hat{z}},k_{+\hat{x}}$&$c_{+\hat{x}-\hat{z}},h_{+\hat{y}-\hat{z}}$&$\alpha$&$\gamma$\\
\hline
$j$&$b,h,i_{+\hat{z}},k_{+\hat{x}+\hat{z}}$&$a_{+\hat{x}},l_{+\hat{x}}$&$\beta$&$\gamma$\\
\hline
$k$&$a,f,i_{-\hat{x}},j_{-\hat{x}-\hat{z}}$&$e_{-\hat{z}},g$&$\gamma$&$\gamma$\\
\hline
$l$&$b,c,e,g$&$a,j_{-\hat{x}}$&$\gamma$&$\gamma$\\
\hline
\end{tabular}
\caption{\label{TableII}
Connections and ground state candidates of the classical Heisenberg model 
on the distorted windmill ($\beta$-Mn) lattice. When the lattice parameter, $y=\frac{9-\sqrt{33}}{16}$, 
then the nearest neighbors 1 and 2 are at equivalent distances. The final column shows the two spin structures 
minimization finds.  Here $\{\alpha,\beta,\gamma\}$ refer to 120$^\circ$ rotated spins.  
In moving to the next unit cell, the $s_{q=0}$ structure is unchanged, while the
$s_{q=2\pi/3}$ structure
replaces $\alpha\rightarrow\beta\rightarrow\gamma\rightarrow\alpha$. }
\end{table}

 Let us refer to spin configurations which order with ordering wavevectors, 
$\mathbf{Q}_\text{0}$ and  $\mathbf{Q}_\text{sp}$, as $\mathbf{Q}_\text{0}$ and $\mathbf{Q}_\text{sp}$ states respectively.
To understand the selection of the $\mathbf{Q}_\text{0}$ state
over other states, we carried out MC simulations  
with a $\mathbf{Q}_\text{sp}$  state 
and a $\mathbf{Q}_\text{mix}$ state that has three planes with $s_{q=0}$ spin configurations and three
planes with $s_{q=2\pi/3}$ spin configurations for $L=6$
as initial states, which allows the system to remain in these states.
We found that the energy of the $\mathbf{Q}_\text{sp}$ state
just below the transition temperature is lower than the energy of
the $\mathbf{Q}_\text{mix}$ state, and the energy of the $\mathbf{Q}_\text{mix}$ state
is lower than the energy of $\mathbf{Q}_\text{0}$, as shown in
Fig.~\ref{energy_0vssp} (a).
However, the latter state is always selected as a ground state if an initial state has a random configuration.
This implies that the selection between $\mathbf{Q}_\text{0}$ and other states is due to entropy. To confirm
our intuition, we computed the specific heats of the $\mathbf{Q}_\text{0}$,
$\mathbf{Q}_\text{mix}$, and $\mathbf{Q}_\text{sp}$ states, and found that the specific heat of $\mathbf{Q}_\text{0}$ state is larger than that of $\mathbf{Q}_\text{mix}$
or $\mathbf{Q}_\text{sp}$ states below the transition temperature as shown in Fig.~\ref{energy_0vssp} (b).
This confirms there are more low lying modes for the $\mathbf{Q}_\text{0}$ state, which leads to the  selection of the  $\mathbf{Q}_\text{0}$
state over the other states of the degenerate ground state manifold.

\begin{figure}
\includegraphics[scale=0.4]{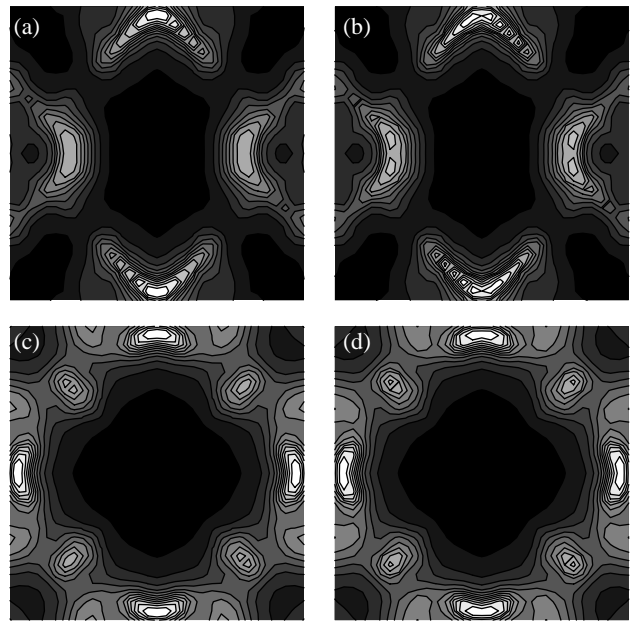}
\caption{\label{figure4}  Contour plots of the intensity of the structure factor in the hhk ((a) and (b)) and hk0 ((c) and (d)) planes for the $\beta$-Mn lattice in the cooperative paramagnetic regime (for the lattice parameter $y=(9-\sqrt{33})/16$, $L=8$ and $T=0.2J$). Classical Monte Carlo ((a) and (c)) agrees well with large-$N$ results ((b) and (d)). The maximal intensity is shown in white. Axes run from $(-4\pi,4\pi)$ in $k$ and $h$, where $k(001)$ describes the vertical axis of (a) and (b), and $h(110)$ the horizontal; $k(010)$ describes the vertical axis of (c) and (d) and $h(100)$ the horizontal.}
\end{figure}

\begin{figure}
\includegraphics[width=3in]{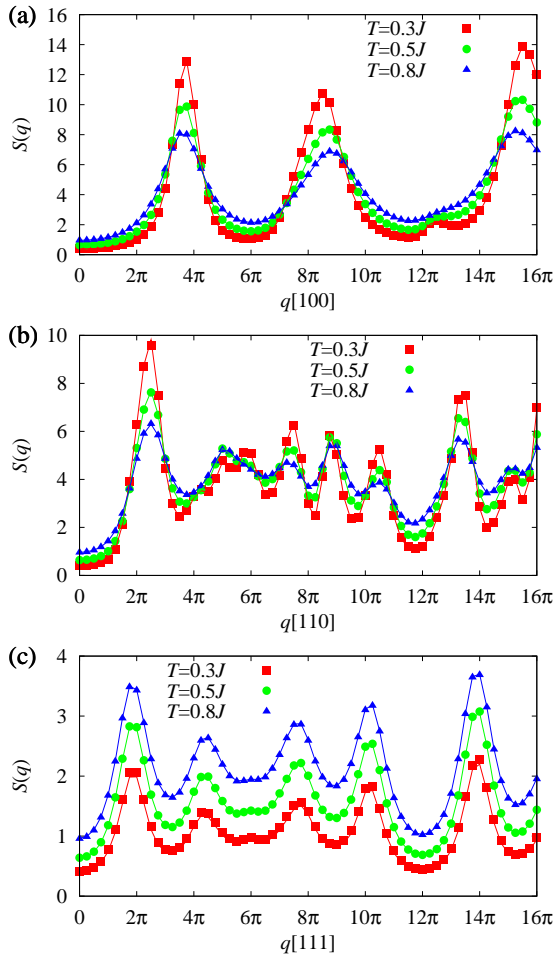}
\caption{\label{figure7} (Color online) A quantitative comparison of the angle-dependent structure factor is shown along three high symmetry directions between the Monte Carlo (Heisenberg model) and large-$N$ results for classical spins in the cooperative paramagnetic phase on $\beta$-Mn lattice ($L=8$). The solid line is the large-$N$  result, while the symbols are from MC simulations.}
\end{figure}

Now let us study the cooperative paramagnetic state above the transition temperature.
Magnetic susceptibility (not shown) has been calculated within both approaches and shows good agreement until very close to the ordering temperature $T_c\approx0.155J$, well below the Curie-Weiss temperature $\Theta_\text{CW}=2J$, and in the zero temperature limit.
Qualitative comparisons between the large-$N$ theory and Monte Carlo results for the structure factor are shown in Fig.~\ref{figure4} in the cooperative paramagnetic phase at temperatures slightly above the onset of magnetic order.  We see excellent qualitative agreement between these two approaches. On the distorted windmill lattice the structure factor intensity is concentrated along the degenerate lines of the large-$N$ result.
Within both approaches 
there is very little weight in the first Brillouin zone due to  the geometric factor. 
In Fig.~\ref{figure7} a quantitative comparison along several high symmetry directions between these two approaches is presented.  There is extremely good agreement at all wavevectors. 

\begin{figure*}
\includegraphics[scale=0.55]{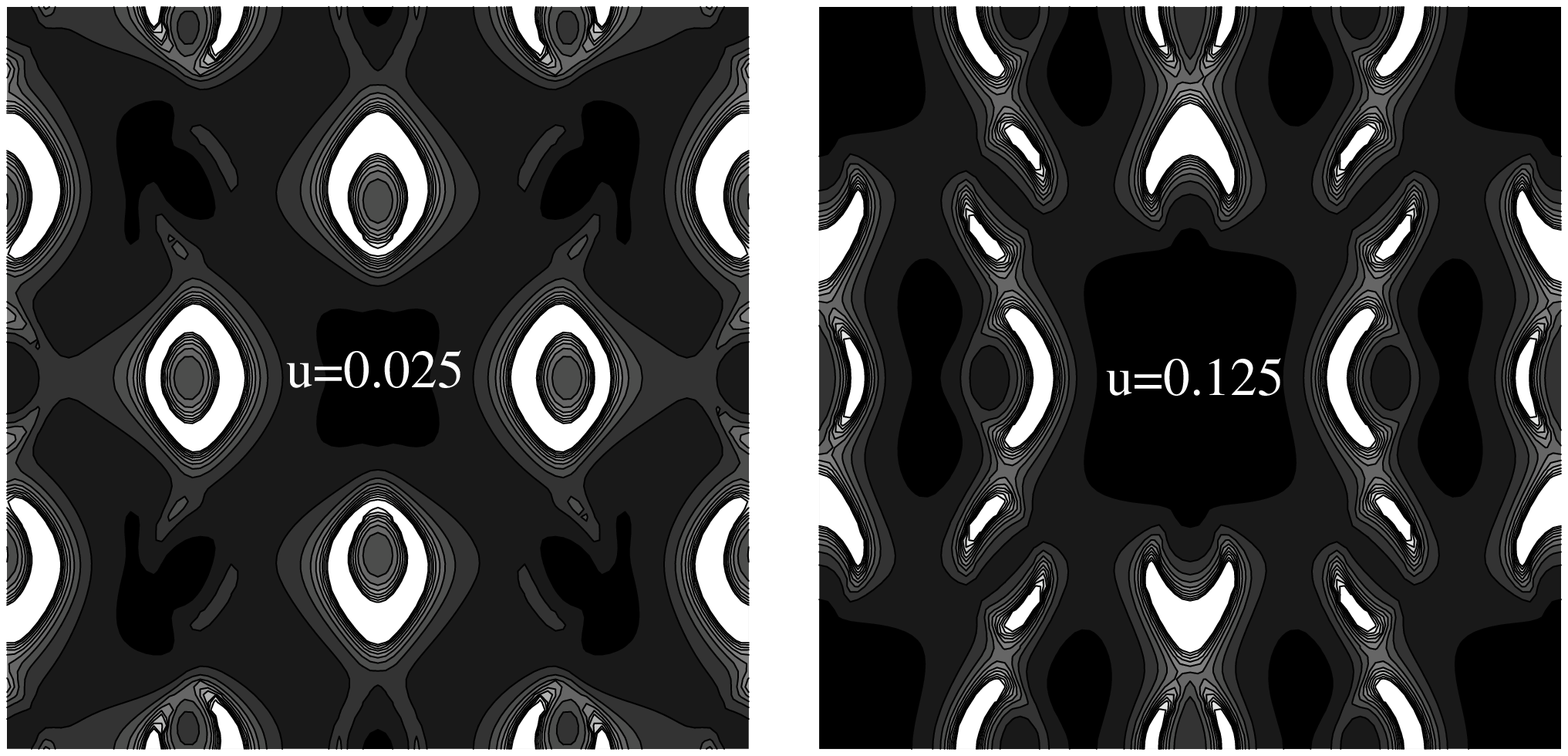}
\caption{\label{figure6}
Contour plots of the intensity of the large-$N$ theory structure factor
in the hhk plane for different values of $u$.  The vertical axis is $k(001)$, and the horizontal axis is $h(110)$, where $k$ and $h$ both run from $(-4\pi,4\pi)$.
Varying the parameter $u$, does not change the energy within either the large-$N$  theory (here $N=3$, $L=24$ and $T=1/4.5 J$), nor in classical Monte Carlo simulations.  It does, however, change the geometric contribution to the structure factor quite dramatically.  While near $u=0.125$ (center) one appears to see few vestiges of the degenerate spheres of the  treatment, for both large (right) and small (left) values of $u$, one sees that these are quite well captured.  Curiously, most real systems seem to lie closer to the middle graph.}
\end{figure*}

\section{ Discussion and summary }

A magnetic material where the magnetic properties can be described by the AF Heisenberg model on the trillium lattice 
is CeIrSi.  
It shows\cite{heying} a Curie-Weiss susceptibility  above 100 K with a magnetic moment of 2.56$\mu_B$/Ce atom 
and a $\theta_{CW}=$ -24 K.  At low temperatures, $\chi^{-1}$ shows a gradual downturn,\cite{heying,note3}
a characteristic common to many triangle-based frustrated magnets.  
 X-ray scattering shows the lattice structure of this material to have a cubic P2$_1$3 symmetry\cite{heying}, 
with a positional parameter $x_{Ce}$=0.6183.\cite{note4}  
Then the Ce sites, which hold magnetic moments consistent with $4f^1$ electrons, 
form a trillium lattice of corner-shared triangles.  The (non-zeroed) interaction matrix is identical to that presented in Eq. 5 
of Ref.~\onlinecite{hopkinson} if one takes the lattice parameter $u_{Ce}$=$x_{Ce}-\frac{1}{2}$.\cite{note2}  
Curiously the energetics of the AF Heisenberg 
model do not depend on the lattice parameter $u$ directly.  As shown in Fig.~\ref{figure6}, one expects that the 
realization of this model with different values of $u$ will again access a cooperative 
paramagnetic regime, but with differing weights over the surface of the MF degenerate spheres.  

To our knowledge Ref.~\onlinecite{heying} is the only study of the physics
of CeIrSi. In this work the magnetic susceptibility and x-ray spectra
of polycrystalline powdered samples have been measured, giving evidence that
this material remains disordered to low temperatures.  
It would be very interesting to see whether or not  neutron scattering
measurement on a single crystal of CeIrSi  may show evidence of partial order
at low temperatures giving way to long range order, as we predict for
an antiferromagnetic Heisenberg model on the trillium lattice.



The relevance of our study to $\beta$-Mn and MnSi is less obvious, as both materials are metallic.  However, the magnetic sites of each lattice features one of the three-dimensional corner-shared triangle lattices structures here studied, with $\beta$-Mn being the only known material to form in the distorted windmill structure.  While the origin of the unusual non-Fermi liquid resistivity seen under pressure in MnSi\cite{doironpedra} ($\Delta \rho \approx AT^{\frac{3}{2}}$) is not at present understood, the same temperature exponent is observed\cite{stewart} in $\beta$-Mn, where it is expected to result from antiferromagnetic spin fluctuations.  We believe it is worthwhile to investigate the possible link between the magnetic fluctuations and the non-Fermi liquid behavior in these materials.  In this vein, it is interesting to note that powder neutron scattering down to 1.4 K\cite{nakamura} in $\beta$-Mn shows no signature of magnetic order.  We are hopeful that this study might provide the motivation for single crystal neutron scattering to be carried out on $\beta$-Mn.

In summary, we have used large-$N$ theory for O($N$) vector spins and classical MC simulations to study the AF Heisenberg model on two three-dimensional corner-shared triangle lattices, each site of which belongs to 
three equilateral triangles.  
The large-$N$ studies suggested that the geometrical frustration present would lead to a partially ordered state on both lattices.
However, through the minimization of finite size spin clusters, we found the ground state  manifolds on these two lattices to be quite different, 
despite the local similarity between these corner sharing triangle lattice structures. 
In both cases, we found that there is a first order transition to a magnetically ordered state using MC methods.
We further showed that the trillium lattice exhibits a unique ground state with a spiral ordering, 
while the distorted windmill has a macroscopic ground state degeneracy. 
Magnetic ordering of the classical AF Heisenberg model on the distorted windmill lattice is therefore seen to arise via an order by disorder mechanism. The degeneracy of this model on the trillium lattice is seen to be an artificial effect of the soft-constraint of the
large-$N$ theory.

Despite the above noted differences at low temperatures between large-$N$ and MC results,
in the cooperative paramagnetic phase above the transition temperature, 
we find a remarkable resemblance between the respective spin-spin correlations.
This leads us to ask whether the salient features of the large-$N$ theory, the angular and directional dependences of the spin-spin correlations found in
the partially ordered state obtained by large-$N$ theory are present at finite temperatures above the transition temperature.  
As true partial order exhibits long range order along  particular directions only 
as $T \rightarrow 0$, it is not possible to have partial order at any finite temperatures,
since the spin-spin correlation decay exponentially at finite temperatures. 
This being said, the qualitative directional dependence characteristic of a partially ordered state survives above the transition temperature
(note that  the ground state  is smoothly connected to the cooperative paramagnetic phase in large-$N$ theory)
allowing us to conclude that a ``disguised" partial order has been recovered in the cooperative paramagnetic phase.

\section*{Acknowledgments}

This work is supported by NSERC of Canada, Canada Research Chair, the
Canadian Institute for Advanced Research (J. M. H., S. V. I., H. Y. K.),
and the Swiss National Science Foundation (S. V. I.).

\appendix
\section{Degeneracy of the distorted windmill lattice ground state}

To estimate the degeneracy of the ground state, we realize that fixing $\{\alpha, \beta, \gamma\}$ in Table \ref{TableI} leaves us with six tilings of the spins in the unit cell, which we might label $\{s_{q=0}^{1},s_{q=0}^{2},s_{q=0}^{3},s_{q=2\pi/3}^{1},s_{q=2\pi/3}^{2},s_{q=2\pi/3}^{3}\}$. Here $s^{1}$ is taken to be the spin structure as presented in Table \ref{TableI}, $s^{2}$ the same taking $\alpha\rightarrow\beta\rightarrow\gamma\rightarrow\alpha$, and $s^{3}$ the same taking  $\alpha\rightarrow\gamma\rightarrow\beta\rightarrow\alpha$.  In progressing from one unit cell to the next along the ($111$) direction, we can follow $s_{q=0}^{1}$ with either $s_{q=0}^{1}$ or $s_{q=2\pi/3}^{1}$, $s_{q=2\pi/3}^{1}$ can be followed by either $s_{q=2\pi/3}^{2}$ or $s_{q=0}^{2}$.  Replacing $1\rightarrow 2\rightarrow 3\rightarrow 1$ gives the general rules for allowed spin structures (provided the number of $s_{q=2\pi/3}$ structures is a multiple of three).  With these rules, it becomes a problem in combinatorics to determine the ground state degeneracy.

For simplicity,\cite{notefinite} let us consider a finite size lattice with $L\times L\times L$ unit cells, with $L$ divisible by 3.  Clearly we have one state with only $s_{q=2\pi/3}^{i}$ spin structures.  By removing three $s_{q=2\pi/3}^{i}$ spin structures, we can insert three $s_{q=0}^{j}$ spin structures in their place, or more generally, removing $3m$ planes of unit cells which rotate as they progress along the ($111$) direction, allows the introduction of $3m$ planes of spin structures which need not rotate from one plane to the next.  Upon such a removal there should be $3l-3m$ available unit cell planes for $3m$ equivalent spin structures.  This implies a degeneracy of approximately\cite{approx2}
\begin{equation}
N_{lm}=\frac{(3l-1)!}{(3m)!(3l-3m-1)!},
\end{equation}
where $m$ can be $\{0,1,..,l\}$.  Summing over $m$ then gives the total number of states:
\begin{eqnarray}
N_L&\approx&\sum_{m=1}^{l-1}\frac{(3l-1)!}{(3m)!(3l-3m-1)!}\\&\approx&\frac{1}{3}\sum_{m=0}^{3l-1}\frac{(3l-1)!}{m!(3l-m-1)!}=\frac{2^{3l-1}}{3}=\frac{2^L}{6},
\end{eqnarray}
so the number of ground states grows exponentially in the linear lattice size, roughly as $N_L\approx \frac{e^{\ln2 L}}{6}\approx \frac{e^{0.69L}}{6}$.



\begin{thebibliography}{00}

\bibitem{reimersmc}
J. N. Reimers,
\prb {\bf 45}, 7287, (1992).

\bibitem{moessenrchalker}
R. Moessner and J. T. Chalker,
\prl {\bf 80}, 2929 (1998);
\prb {\bf 58}, 12049 (1998).

\bibitem{hopkinson} J. M. Hopkinson and H.-Y. Kee, Phys. Rev. B {\bf{74}}, 224441 (2006).

\bibitem{canals} B. Canals and C. Lacroix, Phys. Rev. B {\bf{61}}, 11251 (2000).

\bibitem{bergman} D. Bergman, J. Alicea, E. Gull, S. Trebst and L. Balents, Nature Physics {\bf{3}}, 487 (2007).

\bibitem{unpublishedus} J. M. Hopkinson and H.-Y. Kee, unpublished.

\bibitem{Pflei} C. Pfleiderer, D. Reznik, L. Pintschovius, H. v. L{\"{o}}hneysen, M. Garst and A. Rosch, Nature {\bf{427}}, 227 (2004).

\bibitem{Schroder} A. Schr{\"{o}}der, G. Aeppli, R. Coldea, M. Adams, O. Stockert, H. v. L{\"{o}}hneysen, E. Bucher, R. Ramazashivili and P. Coleman, Nature {\bf{407}}, 351 (2000).

\bibitem{noteref3} This question has been answered in the affirmative for the $J_1$, $J_2$ Heisenberg model on the diamond lattice in Ref.~\onlinecite{bergman}

\bibitem{note1a} True long range order is not captured within the large-$N$ mean field description.

\bibitem{berlin}
T. H. Berlin and M. Kac, Phys. Rev. \textbf{86}, 821 (1952).

\bibitem{jmol} Created by Jmol: an open-source Java viewer for chemical structures in 3D. http://www.jmol.org/ 

\bibitem{nakamura1} See ``Site II'' in Table II of Ref.~\onlinecite{nakamura}.

\bibitem{stanley}
H. E. Stanley, Phys. Rev. \textbf{176}, 718 (1968).

\bibitem{garcan}
D. A. Garanin and B. Canals, \prb {\bf59}, 443 (1999).

\bibitem{isakov}
S. V. Isakov, K. Gregor, R. Moessner, S. L. Sondhi,
\prl {\bf 93}, 167204 (2004).

\bibitem{prv} Note that in Ref.\onlinecite{hopkinson}, the spins were normalized to 1.

\bibitem{andothers} Or along a lattice vector related to this by the symmetry of the crystal structure: $(\pm\frac{2\pi}{3},0,0)$, $(0,\pm \frac{2\pi}{3},0)$ or $(0,0,\pm\frac{2\pi}{3})$.

\bibitem{specifically} As the structure factor geometrically vanishes at
the ordering wavevector in the first Brillouin zone, we define the order
parameter in terms of the structure factors at equivalent wavevectors
$\mathbf{Q} = (\frac{8\pi}{3},0,0)$ and $\mathbf{Q} = (4\pi,0,0)$
for the trillium and distorted windmill lattices respectively.


\bibitem{lawler} M. Lawler, H.-Y. Kee, Y. B. Kim, and A. Vishwanath, arXiv:0705.0990.

\bibitem{note} Large-$N$ mean field theory replaces the local spin constraint, $S_i^2=1$, with a global spherical approximation constraint,\cite{berlin} $\sum_{i=1}^{n_s} S_i^2 = n_s$, where $n_s$ is the number of sites.

\bibitem{usprl} J. M. Hopkinson, S. V. Isakov, H.-Y. Kee and Y. B. Kim, Phys. Rev. Lett. {\bf{99}}, 037201 (2007).


\bibitem{doironpedra} N. Doiron-Leyraud {\it{et al.}}, Nature {\bf{425}}, 595 (2003); P. Pedrazzini {\it{et al.}}, Physica B {\bf{378-380}}, 165 (2006).
\bibitem{nakamura} H. Nakamura, K. Yoshimoto, M. Shiga, M. Nishi and K. Kakurai, J. Phys.: Cond. Matt. {\bf{9}} 4701 (1997).

\bibitem{stewart} J. R. Stewart, B. D. Rainford, R. S. Eccleston, and R. Cywinski, Phys. Rev. Lett. {\bf{89}}, 186403 (2002)

\bibitem{detail} It is possible that not all of the equilateral corner-shared triangles of $\beta$-Mn are of exactly the same size.  Here we assume the structural parameter $y_0=(9-\sqrt{33})/16$ for which they are, which appears consistent with scatter in experimental values as noted in Ref.~\onlinecite{canals}
\bibitem{hopkinson2} J. M. Hopkinson and H.-Y. Kee, Phys. Rev. B {\bf{75}}, 064430 (2007).

\bibitem{heying} B. Heying, R. P{\"{o}}ttgen, M. Valldor, U. Ch. Rodewald, R. Mishra, and R.-D. Hoffman, Monatshefte f{\"{u}}r Chemie {\bf{135}}, 1335 (2004).
\bibitem{note3} 
It is interesting to note that such a downturn is captured in Monte Carlo solutions of the classical Heisenberg model--becoming more pronounced as $N$ decreases towards $N=1$ corresponding to Ising spins;
S. V. Isakov, J. M. Hopkinson and H.-Y. Kee, unpublished.

\bibitem{note4} The positional parameter corresponds to the location within the first unit cell with coordinates ($x,x,x$). The other three Ce atoms within the first unit cell should then be given by $\{x+\frac{1}{2},\frac{3}{2}-x,1-x\}$, $\{1-x,x+\frac{1}{2},\frac{3}{2}-x\}$, and
 $\{\frac{3}{2}-x,1-x,x+\frac{1}{2}\}$.

\bibitem{note2} In the graphs here we have used $u=0.138$ which is the value relevant for the Mn sites on the MnSi lattice.  This corresponds to the distance $\sqrt{(\frac{1}{2})^2+(\frac{1}{2}-2u)^2+(2u)^2}$. Taking $u_{Ce}=0.1183$ describes the nearest neighbour distances between Ce spins since $\sqrt{(\frac{1}{2})^2+(\frac{3}{2}-2u)^2+(1-2u)^2}=\sqrt{(\frac{1}{2})^2+(\frac{1}{2}-2(x-\frac{1}{2})^2+(2(x-\frac{1}{2}))^2}$.

\bibitem{notefinite} In the thermodynamic limit corrections resulting from this assumption will be small.

\bibitem{approx2} Some of these states may be equivalent, but we expect this to contribute at most a factor of $\frac{1}{3l-3m}$, which should merely be a logarithmic correction to our result.

\end{thebibliography}
\end{document}